\documentclass[aps,prx,preprint,superscriptaddress,groupedaddress,amssymb,amsmath,floatfix]{revtex4-1} 
\usepackage{graphicx}  
\usepackage{dcolumn}   
\usepackage{bm}        
\usepackage[breaklinks=true,urlcolor=blue,bookmarks=true]{hyperref} 
\usepackage{color}
\usepackage{upgreek}

\begin{document}



\title{\Large{Emergence of micro-frequency comb via limit cycles in dissipatively coupled condensates}}

\author{Seonghoon Kim$^1$}\email{shoon@umich.edu}
\author{ Yuri G. Rubo$^{2}$, Timothy C. H. Liew$^{3}$, Sebastian Brodbeck$^4$, Christian Schneider$^4$, Sven H\"ofling$^{4,5}$}
\author{Hui~Deng$^{6}$} \email{dengh@umich.edu}
\small{ }
\address{$^1$ Department of Electrical Engineering and Computer Science, University of Michigan, Ann Arbor, MI 48109, USA}
\address{$^2$Instituto de Energ\'ias Renovables, Universidad Nacional Aut\'onoma de M\'exico, Temixco, Morelos 62580, Mexico}
\address{$^3$Division of Physics and Applied Physics, School of Physical and Mathematical Sciences, Nanyang Technological University, 21 Nanyang Link, Singapore 637371, Singapore}
\address{$^4$ Technische Physik, Universit\"at W\"urzburg, Am Hubland, W\"urzburg 97074, Germany}
\address{$^5$ SUPA, School of Physics and Astronomy, University of St Andrews, St Andrews KY16 9SS, United Kingdom}
\address{$^6$ Department of Physics, University of Michigan, 450 Church Street, Ann Arbor, MI 48109, USA}



\begin{abstract}
Self-sustained oscillations, limit cycles, are a fundamental phenomenon unique to nonlinear dynamic systems of high-dimensional phase space. They enable understanding of a wide range of cyclic processes in natural, social and engineering systems. Here we show that limit cycles form in coupled polariton cavities following the breaking of Josephson coupling, leading to frequency-comb emission. The limit cycles and destruction of Josephson coupling both appear due to interplay between strong polariton-polariton interaction and a dissipative contribution to the cavity coupling. The resulting nonlinear dynamics of the condensates is characterized by asymmetric population distribution and nontrivial average phase difference between the two condensates, and by time-periodic modulation of their amplitudes and phases. The latter is manifested by coherent emission of new equidistant frequency components. The emission spectrum resembles that of a micro-frequency comb, but originates from a fundamentally different mechanism than that of existing frequency combs. It allows non-resonant excitation with a power input much below the conventional semiconductor laser threshold. The comb line spacing is determined by the interaction and coupling strengths, and is adjustable up to multi-terahertz frequency. The work establishes coupled polariton cavities as an experimental platform for rich nonlinear dynamic phenomena.
\end{abstract}
\maketitle

In nonlinear dynamic systems with multi-dimensional phase space, self-sustained oscillations, called limit cycles, may emerge from a stable fixed point when the system looses stability through Hopf bifurcation \cite{strogatz_nonlinear_2015}. It is fundamentally different from periodic orbits in linear systems and has no counter part in nonlinear systems of one-dimensional phase space. The oscillation becomes self-sustained; the oscillation frequency is set by the intrinsic dynamical properties of the system rather than initial conditions or the driving frequency. Studies of limit cycles have enabled understanding cyclic phenomena in nonlinear dynamical systems that are ubiquitous in our world, such as the beating of a heart \cite{babloyantz_is_1988}, firing of nerve cells \cite{rinzel_analysis_1998}, chemical oscillations \cite{field_oscillations_1974}, predator-prey interactions \cite{may_limit_1972}, airplane propeller whirls \cite{carroll_bifurcation_1982}, and relaxation oscillations in nonlinear electronic circuits \cite{van_der_pol_forced_1927}. The control of such intrinsic instabilities in nonlinear dynamic systems is not only crucial for many engineering systems but also may enable efficient neuromorphic computing \cite{bertschinger_real-time_2004,opala_neuromorphic_2018} and the simulation of many-body phase transitions \cite{foss-feig_emergent_2017}.
Here we demonstrate the emergence of limit cycles in a pair of coupled semiconductor exciton-polariton (EP) condensates, paving the way for creation and control of nonlinear dynamic phenomena in coupled, dissipative many-body systems.

Semiconductor microcavity EPs
are formed by strong coupling between excitons and photons in high quality cavities \cite{weisbuch_observation_1992}. They combine strong excitonic nonlinearity and robust long-range coherence, providing a fertile ground for complex nonlinear wave phenomena \cite{deng_exciton-polariton_2010, carusotto_quantum_2013}. In single cavities, spin switching \cite{amo_excitonpolariton_2010}, critical slowing down \cite{fink_signatures_2018}, solitons \cite{sich_observation_2012}, exceptional points \cite{gao_observation_2015}, as well as multi-mode lasing and beating among modes \cite{tosi_sculpting_2012} have all been observed. With recent developments in cavity engineering \cite{schneider_exciton-polariton_2017}, we can now create multiple, coupled EP sites, 
where many phenomena that result from on-site interactions and coherent Josephson coupling have also been observed, such as Josephson oscillation \cite{abbarchi_macroscopic_2013}, dynamical squeezing \cite{adiyatullin_periodic_2017}, and phase coupling \cite{ohadi_nontrivial_2016}. While these phenomena share similarities to other coupled nonlinear matter-wave systems, microcavity EPs are special in that they are an intrinsically open, driven system. As a result, in addition to the coherent Josephson coupling, EP condensates may also couple dissipatively \cite{aleiner_radiative_2012,rayanov_frequency_2015} when coupling modifies the radiation loss rates of the states. Such dissipative coupling has led to weak-lasing in an EP lattice through a pitchfork bifurcation transition \cite{zhang_weak_2015}, a phenomenon fundamental to nonlinear dynamic systems of one-dimensional phase space. Richer phenomena that emerge in multi-dimensional phase space have also been predicted, including limit cycles and Hopf bifurcation \cite{rayanov_frequency_2015}, which have not been reported in experiments to date.

Here, using a pair of tightly confined, single-mode EP cavities with controlled couplings, we experimentally demonstrate limit cycle oscillations, connected to the commonly-observed stable EP lasing via the Hopf bifurcation. The limit cycle corresponds to a time-periodic modulation of the amplitudes and phases of the EP condensate. It is a uniquely nonlinear-dynamic phenomena and fundamentally different from population oscillation or beating between multiple stable eigen-modes of the system \cite{tosi_sculpting_2012}. It leads to asymmetric population distribution and non-trivial phase relation between the two cavities. Most interestingly, it directly leads to the emergence of equidistant frequency components, in addition to and distinct from any of the static modes defined by the cavities \cite{rayanov_frequency_2015,khan_frequency_2018}. This mechanism of frequency comb generation is fundamentally different from the widely studied ones using micro-resonators or quantum cascade lasers. The latter are based on resonantly-driven cascaded four wave-mixing, and the comb frequencies are determined by the cavity modes. The comb-like EP emission due to limit cycles has a tera-hertz line spacing determined by the nonlinear interaction and coupling strengths and may enable a non-resonantly pumped, low-power source of micro-frequency combs or tera-hertz waves.


To create the limit cycle state, we use a pair of EP condensates trapped close to each other. Photon tunneling between them leads to Josephson coupling, resulting in the formation of bonding (B) and anti-bonding (A) states with split energies. Correspondingly, the radiation loss of the system is also modified, leading to different dissipation rates of the coupled states, which can be described as dissipative coupling between the condensates. The dissipation rate is higher (lower) when the two condensates are in-phase (out of phase) and emission from the condensates interfere constructively (destructively). Consequently, the total coupling becomes non-Hermitian. The interplay between EP interactions and the non-Hermitian coupling between the condensates leads to rich nonlinear dynamic phenomena. 

The dynamics of the system can be described by the driven-dissipative coupled EP equation \cite{rayanov_frequency_2015, ohadi_spontaneous_2015}:
\begin{equation}
\label{eq:CLE}
\frac{d\psi_{L,R}}{dt} = \frac{1}{2}(p_{L,R}\psi_{L,R}-\gamma\psi_{R,L}-\mu{\lvert \psi_{L,R} \rvert}^2\psi_{L,R})-\frac{i}{2}(2\omega_{L,R}\psi_{L,R}-J\psi_{R,L}+\alpha{\lvert \psi_{L,R} \rvert}^2\psi_{L,R}).
\end{equation}
Here $\psi_{L,R}$ is the order parameter of the condensate in each site; $\omega_{L,R}$ are the frequencies of the uncoupled cavity modes; $p_{L,R} = P_{L,R}-\Gamma$ where $P_{L,R}$ is the incoherent pump strength acting on site L,R and $\Gamma$ is the cavity decay rate; $\gamma$ is the dissipative coupling strength; $\mu$ is the gain saturation parameter; $J$ is the Josephson coupling strength; and $\alpha$ is the on-site interaction strength.

\begin{figure*}[t]
	\includegraphics{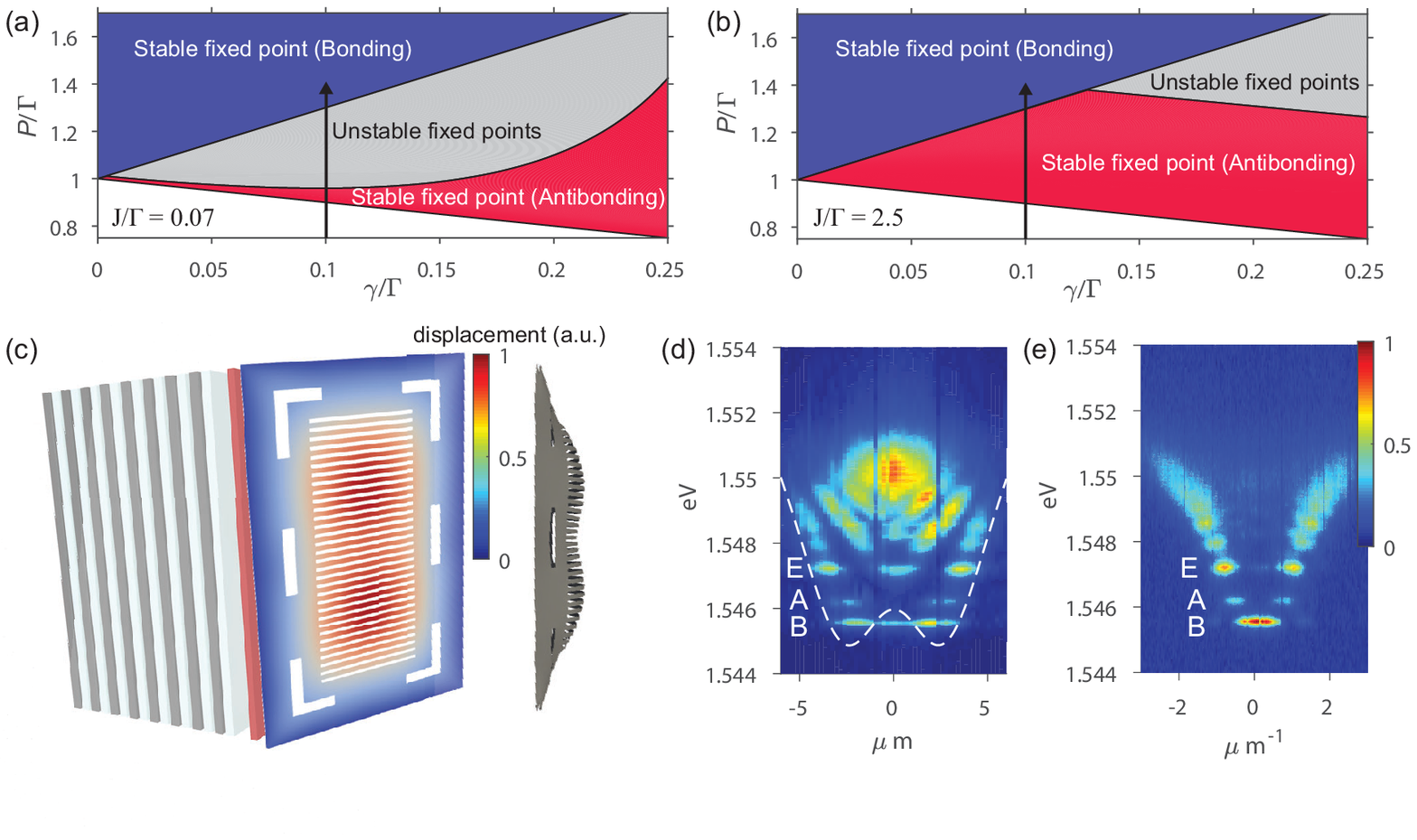}
	\caption{Bifurcation diagram and the sample properties at low excitation powers. (a),(b), Bifrucation diagrams of equation \eqref{eq:CLE} in $P-\gamma$ parameter space for $J/\Gamma = 0.07$, $2.5$, respectively. For both diagrams, $\alpha/\Gamma = 0.25$, $\mu/\Gamma = 0.05$. A standard lasing threshold in the absence of dissipative coupling is at $P=\Gamma$ . For certain values of $\gamma>0$, thresholds for stable and unstable fixed-point solutions emerge as $P$ increases, indicated by arrows for $\gamma/\Gamma = 0.1$. (c), A schematic of the sample structure with a bent SWG mirror. Bending of the SWG is simulated by COMSOL and shown both by the color map in the schematic view and in the side view. The bending is less where there are open slots in the tethering pattern and vice versa. (d), The real-space photoluminescence (PL) spectrum of the coupled polariton system showing the discrete polariton states at low excitation powers. The ground state is formed by the bonding state (B state), while the first excited state is the anti-bonding state (A state). The white dashed line illustrates the potential due to the bending of the SWG shown in (c). (e), The corresponding Fourier space spectrum showing the B state at $k = 0$ and the A state at $k = \pm\pi/a$, where $a$ is the distance between the two coupled sites.}
	\label{fig:s}
\end{figure*}

Two examples of phase diagrams based on Eq.~\ref{eq:CLE} are shown in Fig.~\ref{fig:s}(a)-(b) as a function of the pump rate $P$ and dissipative coupling rate $\gamma$, for fixed $J/\Gamma= 0.07$ and $2.5$ respectively. With increasing pump rate $P$, the system transitions from a thermal state to two possible types of stable lasing states: weak lasing in a state with the lowest decay rate and lasing in the ground state -- the bonding state of the coupled EP system. Both are fixed point solutions of the coupled system. With a sufficiently large dissipative coupling strength $\gamma$, limit cycle solutions, or comb states, can exist between the two stable fixed points. We note that to produce the limit cycles, a large on-site interaction $\alpha$ is necessary, which can be realized with tight confinement of the EPs \cite{kim_coherent_2016} and is  about $\alpha/\Gamma = 0.25$ in our system. As shown in the figures, for larger $J/\Gamma$ (Fig.~\ref{fig:s}(b)), the unstable fixed-point domain in the parameter space becomes small and requires large dissipative coupling $\gamma$, making it harder to realize experimentally. However, decreasing $J$, coherent coupling between the site, would often lead to a decrease in $\gamma$, the dissipative coupling between the two site. For fixed $J$ and $\gamma$, a relatively large $\Gamma$ may facilitate the creation of the limit cycle state, but it also needs to be sufficiently low to allow stimulated scattering and formation of coherent condensates. Therefore optimal values of $\Gamma$, $J$ and $\alpha$ exist to achieve a comb state.

The device we use is illustrated in Fig. \ref{fig:s}(c). The GaAs-based microcavity is constructed with a subwavelength grating (SWG) top mirror of $6\rm \mu m \times 14 \mu m$ in size and a bottom distributed Bragg reflector (DBR). The properties of single SWG-DBR EP systems and polariton lasing in this system have been studied in detail before \cite{zhang_zero-dimensional_2014,kim_coherent_2016}. Lateral confinement of the EPs and full discretization of the EP modes are created and controlled by the lateral size of the SWG mirror \cite{zhang_zero-dimensional_2014}, while multiple, coupled EP sites are created by controlling the placements of the grating bars and the tethering pattern surrounding the SWG \cite{zhang_coupling_2015}. The tethering pattern controls the strain release when the sacrificial layer is removed and in turn controls the bending of the individual grating bars, as shown by simulations using COMSOL Multiphysics (Figure~\ref{fig:s}(c)). The bending of the SWG directly modulates the cavity length and thus the exciton-cavity detuning, through which we form a trapping potential for EPs. Controlling the location and shape of the tethering patterns, therefore, controls both the height and width of the potential barrier between two sites. This tunability provides us control over both the on-site interaction \cite{kim_coherent_2016} and inter-site coupling \cite{zhang_coupling_2015} of the EPs.


Figure \ref{fig:s}(d) and (e) show the real-space and Fourier-space photoluminescence (PL) spectra of the EPs at low pump powers. The states are discretized due to tight confinement, which also enhances the on-site interaction to the order of 10~$\rm \mu$eV due to increased EP density \cite{kim_coherent_2016}.  Bonding (B) and anti-bonding (A) states are formed due to Josephson coupling. Their separation gives the coupling strength $J=0.5$~meV. The Josephson coupling decreases as the pump power increases, because the pump is located at the center of the device and creates a local carrier population that adds to the potential barrier between the two sites (see Supplementary Figure S1 for the narrowing of energy separation between the bonding and anti-bonding states). The distance between the two minima of the effective potential is 6.4~$\rm \mu m$, corresponding to the separation of the two cavity sites. The next lowest energy state, labeled as state E in Figure \ref{fig:s}, is formed from the first excited state of each uncoupled site. It is 1.4~meV above state A, far separated from the lasing frequencies of the stable fixed points or limit cycles.

We first observe signatures of the limit-cycle state through the power dependence of the spectral and spatial distributions of the emission. The power dependence of the real-space spectra is shown in Fig. \ref{fig:pl}. The PL spectrum at the low power (Fig.~\ref{fig:s}(d),(e)) shows clearly the eigen-states of the system, with the three lowest ones the B state, A state and E state. The bonding state B initially has a larger population than the state A, because the pump spot is placed at the center of the device and prefers the B state. As the pump power increases, the anti-bonding state A becomes more populated than state B (Fig. \ref{fig:pl}(a),(b),(f)). This corresponds to the onset of weak lasing at $P=\Gamma-\gamma$ in the state with a lower decay rate\cite{aleiner_radiative_2012}. The apparent asymmetry between the two sites is expected from the symmetry breaking in the weak lasing regime \cite{aleiner_radiative_2012}. With increasing power, limit-cycle oscillations appear, leading to the appearance of new frequency components (Fig. \ref{fig:pl}(c)-(d), (g)-(h)). We fit the spectrum of each site to equidistant Lorentzian lines. Above the bifurcation threshold, up to four equidistant lines are resolved for the right site R (red), and up to three for the left site L (blue), with fitted line spacing of 0.19~meV and 0.27~meV, respectively. We note that all these frequency lines are near the original A and B state and far below the E state. At high pump powers, the PL eventually becomes dominated by the B state and other frequency components become insignificant, showing the system is transitionary toward the single-mode B state lasing (Fig. \ref{fig:pl}(e), (h)). (See supplementary information for the case when single-mode B state lasing is reached in the system.) This second transition takes place at $P=\Gamma+3\gamma$. This sequence of transitions agrees with the dissipative coupling modeled by equation~\eqref{eq:CLE} and allows us to estimate the dissipative coupling rate for our system as $\gamma=0.055ps^{-1}$ (see Appendix for details).


To verify the uniformity of the mode spacing and phase coherence between the multiple frequency lines, we measure the temporal first-order coherence $g^{(1)}(\tau)$. As shown in Fig. \ref{fig:g1}, $g^{(1)}(\tau)$ features a main central peak that decays over a few picoseconds, corresponding to the full spread of the multiple frequency lines. At larger $\tau$, instead of a smooth decay, multiple small peaks are apparent. Such coherence revivals confirm that the emission consists of multiple equidistant, coherent frequency lines.
On the R site, where four frequency components are present, a larger number and more distinct revival peaks are observed compared to the L site. At low power, a clear revival peak is measured at 40~ps for both sites, but other revival peaks are less well resolved due to the low coherence time (Fig. \ref{fig:g1}(a)). As the populations in the two sites grow, the coherence time becomes longer, and the revival peaks become more apparent (Fig. \ref{fig:g1}(b)). At higher powers (Fig. \ref{fig:g1}(c)), the number of revival peaks and the complexity in $g^{(1)}(\tau)$ decrease, suggesting the reduction of frequency components as the system evolves toward the stable fixed point of bonding state lasing. We compare the experimental $g^{(1)}(\tau)$ with computed results based on equation \eqref{eq:CLE} (Fig. \ref{fig:g1}(d),(e),(f), see Appendix for parameter values of the model), which qualitatively captures the main features of the measured coherence revivals.


Lastly, a hallmark of the limit-cycle state is a nontrivial relative phase $\phi$ between the two sites. The dissipative coupling alone favors an out-of-phase relationship between the two sites with $\phi=\pi$. At the same time, on-site interaction changes the instantaneous frequency of each site. Interplay between these two results in a nontrivial phase $\phi\neq 0$ or $\pi$ between the two sites when stable limit-cycle oscillations are formed. We measure the relative phase $\phi$ by interfering both sites simultaneously with a reference beam and fit the interference fringes to obtain their phases \cite{manni_spontaneous_2011}. As shown in figure \ref{fig:rp}, we obtain $\phi=0.51\pm 0.08~\pi$ when the multiple frequency lines appear (Fig. \ref{fig:rp}(a),(d)). When the B state brightens up at high powers, the relative phase changes to $\phi=0.21\pm0.06~\pi$ (Fig. \ref{fig:rp}(b),(e)) and $0.15\pm 0.04~\pi$ (Fig. \ref{fig:rp}(c),(f)), converging toward an in-phase relation for single-mode B state lasing. The above nontrivial phase relationship is also evidenced by the shift of the $k = 0$ peak in the power-dependence of the k-space PL spectra as shown in the supplementary information.


In conclusion, we demonstrate a limit cycle transition through Hopf bifurcation in a pair of dissipatively-coupled nonlinear polariton condensates. Signatures of the limit cycle transition are measured, including the generation of equidistant new frequency components, corresponding coherence revivals in $g^{(1)}(\tau)$, and change of the relative phase between two condensates from $\pi$ toward zero as the system transitions from unstable asymmetric lasing to limit-cycle to stable single-state lasing. These signatures are distinct from those of multi-state lasing, four-wave mixing among the static eigen-modes of the polariton system, or weak lasing. From the excitation dependence of the transition, we infer a dissipative coupling rate of about one tenth of the decay rate. The limit-cycle oscillations lead to the generation of multiple, equidistant lines, resembling a micro-frequency comb. Such a comb allows non-resonant or electrical excitations \cite{suchomel_electrically_2018} with very low input power, as it takes place near the polariton lasing threshold without electronic population inversion. Future work to modify the quality factor and interaction strength of the microcavities may result in narrower linewidth of individual comb lines and greater line spacing. Scaling up the system to a lattice of condensate may provide a platform for efficient neuromorphic computing \cite{bertschinger_real-time_2004,opala_neuromorphic_2018} and simulation of many-body phase transitions \cite{foss-feig_emergent_2017}.

S.K. and H.D. acknowledge the support by the National Science Foundation (NSF) under Awards DMR 1150593 and the Air Force Office of Scientific Research under Awards FA9550-15-1-0240. Y.G.R. acknowledges the support from CONACYT (Mexico) Grant No. 251808. T.L. was supported by the Ministry of Education (Singapore) grant 2017-T2-1-001. C.S., S.B. and S.H. acknowledge the support by State of Bavaria and the Deutsche Forschungsgemeinschaft (DFG) within the project SCHN11376 3-1. The fabrication of the SWG microcavities was performed in the Lurie Nanofabrication Facility (LNF) at Michigan, which is part of the NSF NNIN network.

\section*{\textsc{APPENDIX A: SAMPLE}}

Our microcavity sample is a $\lambda$/2 AlAs cavity with three stacks of four GaAs quantum wells placed at the electric field maxima. The top mirror consists of high contrast subwavelength grating and 2.5 pairs of $\rm Al_{0.15}Ga_{0.85}As/AlAs$ layers and the bottom mirror consists of 30 pairs of $\rm Al_{0.15}Ga_{0.85}As/AlAs$ layers. The Rabi splitting of 12~meV at 5~K was measured from the uncoupled single site. We designed various positions of the release patterns to control interactions between polaritons in the coupled cavities. The sample used in the main text has a center-to-center distance of $6.4~\mu m$ with $6~\mu m$ long grating bars. The length of the grating bar determines the size of the polariton mode in the single site and therefore controls the on-site interaction strengths. The center-to-center distance controls the Josephson and dissipative coupling strengths.
\\

\section*{\textsc{APPENDIX B: MEASUREMENT SETUP}}

The microcavity sample is kept at 5~K in a closed-cycle cryostat. We use a continuous wave Ti:sapphire laser chopped with electo-optic modulator at 5~kHz with 5$\%$ duty cycle. The pump spot is focused on the center of the device with a diameter of 2~$\rm \mu m$. We use a grating-based spectrometer and Michelson interferometer to resolve the spectral lines and beating peaks from these lines.
For the measurement of $\phi$, we use the emission from one of the sites as a phase reference to determine the relative phase between the two sites, as the absolute phase of the polariton condensate is different for every experimental realization. We use a Mach-Zehnder interferometer and magnify the image from one arm by a factor of 6 compared to the other arm. For two spatial modes of $2~\mu m$ in diameter separated by $7~\mu m$, magnification of about 4.5 is needed in order for the single-site to interfere with the entire system. Our magnification ensures that the two sites overlap with the center of the single-site where the phase is uniform. We fit the interference pattern in each site to $I_{L,R}(x) = I_{L,R}(x)(1 + |g^{(1)}_{L,R}|cos(k_{x}x + \phi_{L,R}))$, where $I(x)$ is the Gaussian intensity profile and $k_x$ is the spatial frequency of the fringe pattern due to the angle between two interfering beams. The relative phase is then calculated as $|\phi_{L}-\phi_{R}|$.
\\

\section*{\textsc{APPENDIX C: NUMERICAL SIMULATION}}

We numerically solved equation \ref{eq:CLE} using a fourth-order Adams-Bashforth-Moulton predictor-corrector method with small initial populations in both sites. Note that the initial condition does not affect the final state which converges to the limit-cycle solution. To account for the effect of noise, we multiplied exponential decay functions to the simulated $g^{(1)}(\tau)$. The parameters used for Fig. 3(d),(e),(f) were $\Gamma = 0.5~ps^{-1}$, $\gamma = 0.077~ps^{-1}$, $\omega$ = 0, $J = 0.077~ps^{-1}$, $\alpha = 1.15~ps^{-1}$, and $\mu = 0.015~ps^{-1}$. It is important to note that $\alpha$ in the simulation is the polariton interaction strength multiplied by the polariton population. Considering the polariton population obtained experimentally, one requires the polariton interaction strength to be about 10~$\mu eV$, which is in agreement with previous reports in GaAs polariton systems. We changed the pumping strength $P$ to account for the strength of the excitation power assuming other parameters do not change significantly above the lasing threshold. We used $P = 0.524,~0.548,~0.627~ps^{-1}$ respectively. We also gave a 1$\%$ difference in pumping strength between two sites to account for the asymmetry in experiments.

\begin{figure*}[t]
	\includegraphics[width=\linewidth]{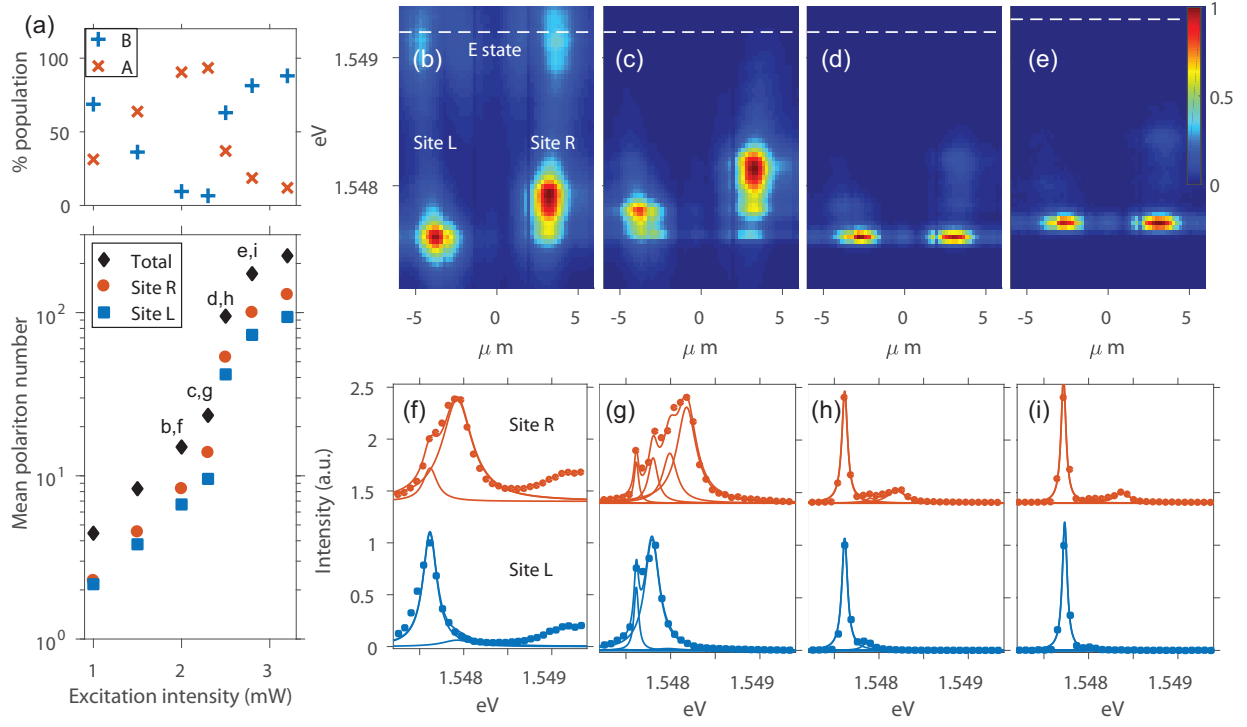}
	\caption{Excitation power dependence of the intensity and real-space spectra of the polariton PL near the A and B states. (a), Bottom: Mean polariton number of the A and B states vs. the excitation power for the L site (blue square), R site (red circle) and the sum of the L and R sites (black diamond). It shows clearly a threshold behavior and degenerate occupation at each site. Top: Relative fraction of the B state (blue plus) and A state (red cross) population vs. the excitation power, showing switching of the dominant state upon transitions to stable weak lasing near A state followed by the limit cycles, and to stable lasing in the B-state. (b)-(e), The real-space spectra at four different excitation powers as marked in (a), showing the transition from weak lasing, to limited cycles, toward B state lasing. The white dashed line marks the E state -- the next lowest energy state above the B and A states. (f)-(i), Spectrum of each site obtained from (b)-(e), respectively. The solid lines are fits by equidistant Lorentzians. More than two frequency components are apparent in (c), (d), (g) \& (h).}
	\label{fig:pl}
\end{figure*}

\begin{figure*}[t]
	\includegraphics[width=\linewidth]{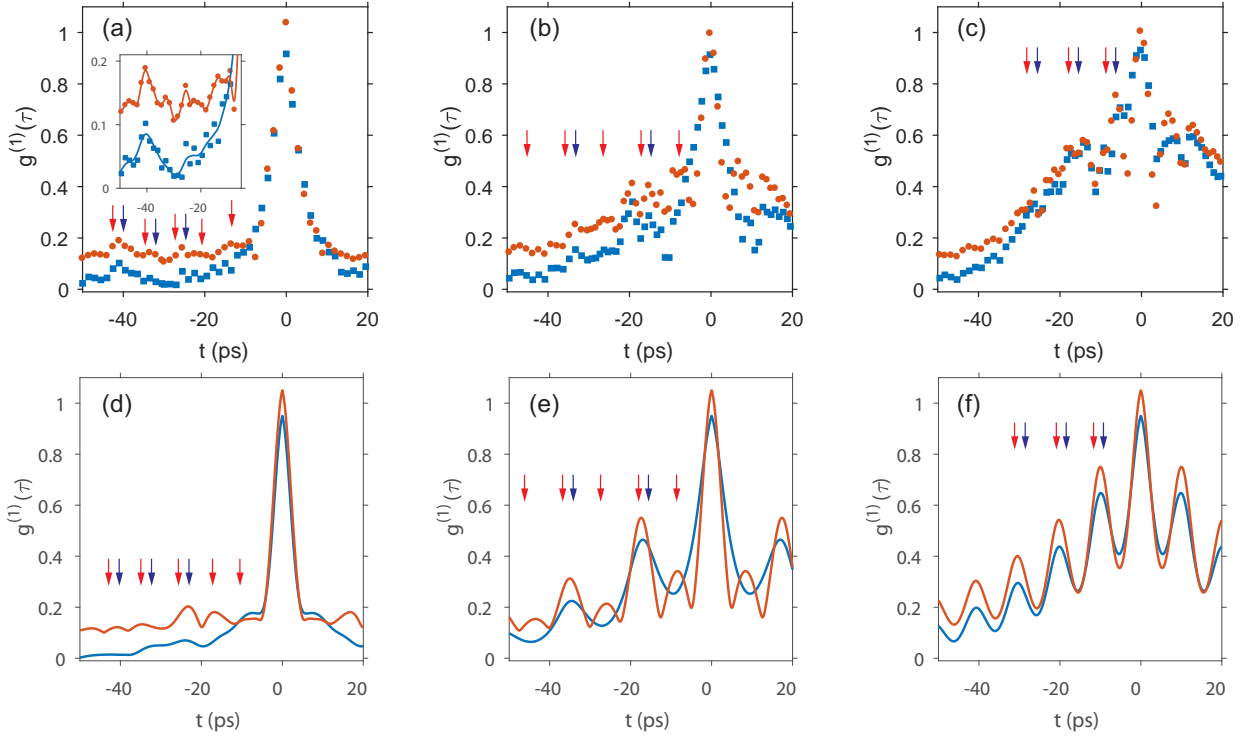}
	\caption{Experimental (a)-(c) and theoretical (d)-(f) $g^{(1)}(\tau)$ for three different excitation powers, corresponding to Fig.~\ref{fig:pl}(c)-(e), respectively. Red dots are shifted vertically by 0.1 for better visibility. The inset shows a zoom in of the revival peaks in (a). The lines are guides to the eye. The blue and red arrows indicate experimentally measured beating peaks and corresponding peaks in simulations for site L and R, respectively. The site with more frequency lines has more revival peaks and narrower $g^{(1)}(\tau)$ linewidths at $\tau=0$. The simulated $g^{(1)}(\tau)$ is multiplied by an exponential decay with decay times of 20, 25, 30~ps respectively.}
	\label{fig:g1}
\end{figure*}

\begin{figure*}[t]
	\includegraphics[width=\linewidth]{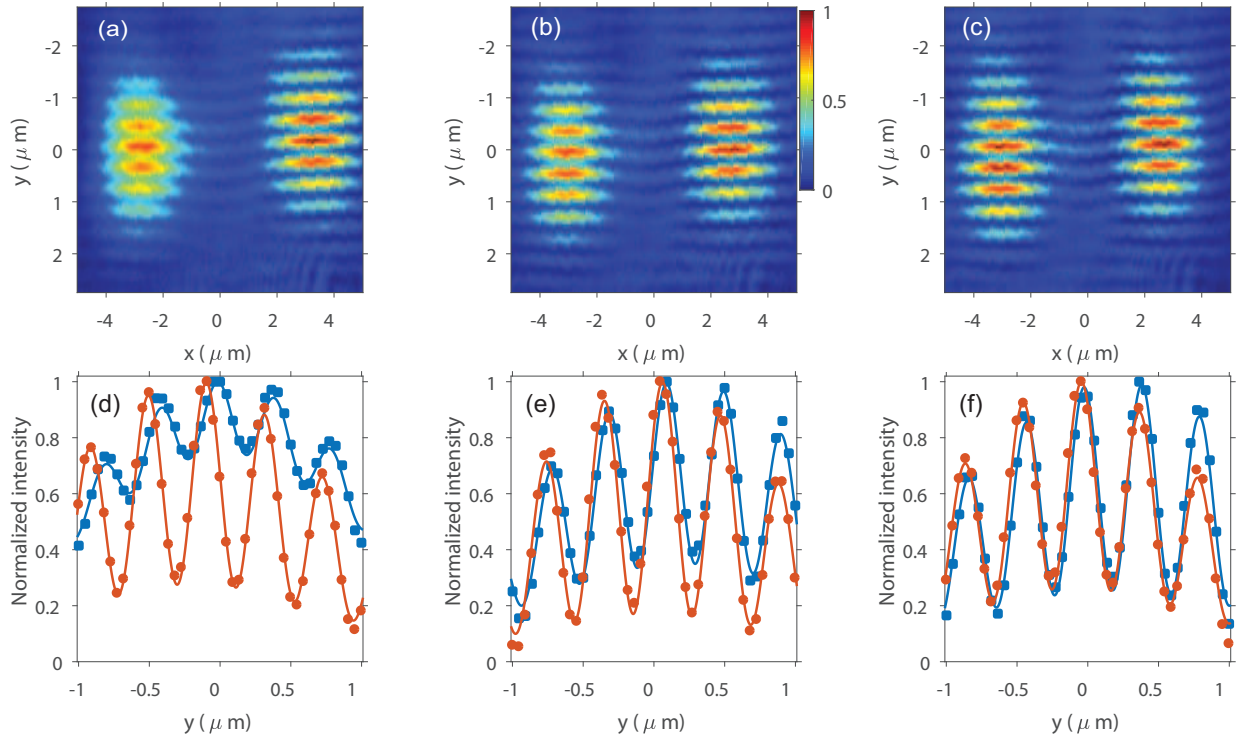}
	\caption{Relative phase measurement between L and R sites at excitation powers of 2~mW (a),(d), 2.3~mW (b),(e) and 2.5~mW (c),(f). (a)-(c), Interference images from interfering both sites to a magnified single site. (d)-(f), Interference fringes for each site obtained from (a)-(c) along $x=\pm3~\mu m$ (dots). The solid lines are fits as described in Appendix. From the fit, we obtain the relative phase difference of $0.51\pm0.08~\pi$, 0.21$\pm0.06~\pi$, and $0.15\pm0.04~\pi$ respectively.}
	\label{fig:rp}
\end{figure*}

The dissipative coupling strength can be estimated by the observed thresholds. It is convenient to express equation \eqref{eq:CLE} based on a pseudospin vector defined as $\mathbf{S}=\frac{1}{2}(\mathbf{\Psi^{\dagger}}\cdot\bm{\sigma}\cdot\mathbf{\Psi})$, where $\mathbf{\Psi}=(\psi_{1},\psi_{2})^{T}$ and $\bm{\sigma}$ is the Pauli vector.
\begin{subequations}
	\begin{align*}
	\frac{dS_x}{dt} &= (p-\mu S)S_x-\gamma S- \alpha S_zS_y
	\\
	\frac{dS_y}{dt} &= (p-\mu S)S_y+JS_z+ \alpha S_zS_x
	\\
	\frac{dS_z}{dt} &= (p-2\mu S)S_z-JS_y
	\\
	\frac{dS}{dt} &= (p-\mu S)S-\mu S_z^2-\gamma S_x
	\end{align*}
\end{subequations}
where $S_x = \frac{1}{2}(\psi_2^*\psi_1+\psi_1^*\psi_2),~S_y = \frac{i}{2}(\psi_2^*\psi_1-\psi_1^*\psi_2),~S_z = \frac{1}{2}(\lvert \psi_1 \rvert^2 - \lvert \psi_2 \rvert^2),~S = \frac{1}{2}(\lvert \psi_1 \rvert^2 + \lvert \psi_2 \rvert^2)$. Then the nontrivial fixed point A state becomes stable  when $p=-\gamma$ and $S_x=-S,~S_y=S_z=0,~S=(\gamma+p)/\mu$. This happened at the pump power of about 2~mW in the experiment (Fig. \ref{fig:pl}(b)). The threshold for the stable fixed point B state is when $p=3\gamma$ and $S_x=S$. This corresponds to the pump power of about 3~mW when the system stabilized to the B state with weak satellite peaks (Fig. \ref{fig:pl}(e)). Assuming $\Gamma = 0.5~ps^{-1}$ and $P$ is a linear function of pump power, the estimated dissipative coupling strength is about $0.055~ps^{-1}$ which is reasonable considering the value we used for the simulation.
\\

%
\pagebreak

\end{document}